\documentclass[aps,prl,reprint,floatfix,amsmath,amssymb,
superscriptaddress,footinbib]{revtex4-1}

\usepackage{lipsum}
\usepackage{graphicx}
\usepackage{epstopdf}
\usepackage[dvipsnames]{xcolor}
\usepackage{hyperref}
\usepackage{bm}
\usepackage{printlen}
\usepackage{units}
\usepackage{braket}

\hypersetup{
        colorlinks,
   citecolor=blue,
   filecolor=blue,
   linkcolor=blue,
   urlcolor=blue,
   breaklinks=true}

\renewcommand{\vec}[1]{\mathbf{#1}}

\newcommand{\Trace}{\text{Tr}\,}

\renewcommand{\Im}{\text{Im}}

\newcommand{\eps}{{ \varepsilon}}

\newcommand{\Ham}{{\mathcal H}}

\newcommand{\PT}{$\mathcal{PT}$}

\hypersetup{hidelinks}

\usepackage{xcolor}

\begin{document}

\title{Anomalous conductivity of   $\mathcal{PT}$-symmetric Fermi liquids}

\author{Alexander Kruchkov}

\affiliation{Institute of Physics, {\'E}cole Polytechnique F{\'e}d{\'e}rale de Lausanne,  Lausanne, CH 1015, Switzerland}

\affiliation{Department of Physics, Harvard University, Cambridge, Massachusetts 02138, USA}  

\affiliation{Branco Weiss Society in Science, ETH Zurich, Zurich, CH 8092, Switzerland}

\begin{abstract}
We consider a non-Hermitian yet $\mathcal{PT}$-symmetric Fermi liquid (\PT-FL) in external electric fields. Due to $\mathcal{PT}$-symmetry, the system exhibits real spectrum,   Fermi surface and electric conductivity are well-defined through propagators. We find that, in contrast to the  conventional  Fermi liquids (FL), the $\mathcal{PT}$-FL can exhibit a \textit{zero resistance state} in the longitudinal ($xx$) channel.   Moreover, the temperature dependence of the resistivity anomaly violates the conventional FL scaling (it is \textit{not} limited by $T^2$).   These findings open route to further exploration of transport anomalies beyond the conventional paradigm. 
\end{abstract}

\maketitle

\textit{Zero-resistance states} are rare beauty in condensed matter physics and are generically associated with the superconducting states. Superconductivity as a macroscopic quantum phenomena is characterized by formation of electron pairs (Cooper pairs), which behave as bosons, and hence can exhibit a superfluid state at low temperatures.  This results into electric currents flowing without dissipation, i.e. experiencing \textit{zero resistance} from the system \cite{Tinkham2004}.
Another famous example is zero resistance  in Quantum Hall Effect (QHE) \cite{Klitzing1980,Tsui1982a, Tsui1982}, for which resistance (conductance) in transverse ($xy$) direction is quantized, while in the longitudinal ($xx$) direction is zero. 
While the two famous examples above refer to DC zero-resistance (the frequency of external fields $\omega$$\to$$0$), it is known that under proper engineering one can tune the AC response to zero resistance \cite{Mani2002,Zudov2003, Willett2004}, i.e. engineer quantum state for which conductivity spikes at finite frequency $\omega_*$. For the theoretical account 
see  e.g. Ref.\cite{Shi2003}.    
Finally, the  question of ideal dissipationless conductivity  had been researched in the context of  one-dimensional (1D) Hamiltonians \cite{Castella1995,Zotos1997}. While that research had established very interesting links between dissipationless currents in 1D and the integrability of the system, it does not seem possible to extend that formalism to generic 2D and 3D currents.

In the Fermi liquid (FL) theory,---a basic theory for electronic conductivity in normal  metals,---quasiparticles are well behaved and  are reminiscent of bare electrons. The electric conductivity of a conventional 3D Fermi liquid is given by 
\begin{align}
    \sigma_{\text{FL}}  = \frac{n e^2 \tau}{m}  
        \label{Drude}
\end{align} 
(here $n$, $m$, $\tau$ are electronic density, mass, and scattering time), is smooth and singularity-free.  This is linked to the smoothness of the quasiparticle spectrum both for non-interacting and interacting FLs,  and smooth, topologically-trivial Fermi surface. Hence, anomalies in behavior of Fermi liquids, and in particular zero-resistance states, are not expected.

On the other hand, a particular interesting topic in novel phases of quantum matter is exploration of non-hermitian systems with real spectra \cite{Ganay2018}.  This can be achieved for the non-hermitian Hamiltonian, which acquire additional symmetry under parity and time-reversal (\PT) \cite{Bender1998,Mostafazadeh2002}.  On the condensed matter side, \PT symmetry, which reverses the signs of time and coordinates simultaneously, plays an important role in quantum transport  and typically leads  to the notions of topological protection \cite{Chiu2016,Armitage2018}.   Experimentally, various \PT-symmetric Hamiltonians have been operated by  e.g.   controlling gain and loss in quantum optics simulators \cite{PTexp1,PTexp2,PTexp3,PTexp4,PTexp5}. It has been demonstrated that a \PT-symmetric 2D crystal can be operated in practice \cite{Kremer2019,Szameit2011}. On the theory side, the reality of the energy spectrum of a non-Hermitian, yet \PT-symmetric condensed matter Hamiltonian assures mapping to an effective  Hermitian Hamiltonian with the associated unitary evolution \cite{Bergholtz2021}, and allows to define Green's functions, Fermi surface, and electric conductivity.

In this Letter, we address aspects of re-engineering  Fermi surface and its topology in 2D systems in the way it  could give singularities to the fermionic conductivity. The Fermi surface re-engineering can be achieved by e.g. lowering the system’s symmetry to \PT symmetry.  We show that in contrast to conventional Fermi liquids \eqref{Drude}, a \PT-symmetric Fermi liquid exhibits anomalies in the electric conductivity, approaching possible zero-resistance phase. The source of this   low resistance/zero resistance anomaly is the non-hermitian perturbation of the Hamiltonian, which re-engineers the topology of the Fermi surface, leading to very large Fermi velocities. Notably, the temperature dependence of resistivity in this phase is shown to deviate from the conventional Fermi liquid law. We argue that this anomaly could be achieved under the existing experimental conditions of the ultracold fermions on the honeycomb lattice \cite{Jotzu2014}, where complex-valued next-nearest-neighbour hoppings can be carefully engineered.

\begin{figure} [t ]
\includegraphics[width = 0.95 \columnwidth]{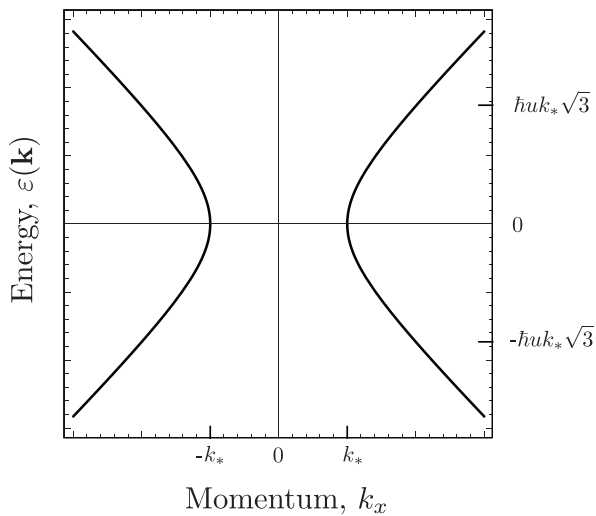}
\caption{The energy spectrum of PT-symmetric Hamiltonian \eqref{Hamiltonian}  \label{spectrum}.}
\end{figure}

\textit{Hamiltonian.}---Without loss of generality we consider a \PT-symmetric Hamiltonian in 2D as
\begin{align}
   \mathcal H = \hbar u \, \boldsymbol{\sigma} \cdot \vec k + i \eta \sigma_z ,
   \label{Hamiltonian}
\end{align}
where $\sigma$'s are Pauli matrices,  $u$ is the system's parameter with dimensionality of velocity (m/s), $\eta$ is the strength of non-hermiticity in the system.  Such a Hamiltonian can be simulated experimentally on the hexagonal lattice with gain and loss \cite{Kremer2019,Jotzu2014}. The Hamiltonian \eqref{Hamiltonian} is explicitly non-Hermitian ($\Ham^{\dag} \ne \Ham$), however it exhibits  \PT symmetry and hence has a real-valued energy spectrum (Fig.\ref{spectrum}). In what follows below, we consider 2D case assuming a potential  links to experiment setups of type \cite{Jotzu2014}; generalization to a 3D case is straightforward. In what follows below we consider fermions as charge carriers, and refer to  them as "electrons" for comparison with conventional Fermi liquids \ref{Drude}; yet they could be Fermi atoms in optical lattices as well.

First, by direct diagonalization of \eqref{Hamiltonian}, we find that the system exhibits gapless spectrum with two fermionic bands 
\begin{align}
 \varepsilon_{1,2} (\vec k) = \pm \hbar u \sqrt{k^2 - k^2_*}, 
 \label{dispersion}
\end{align} 
Note that both electronic bands \eqref{dispersion} has a real-valued electronic spectrum (see Fig.\ref{spectrum}), which is a consequence of \PT symmetry of the Hamiltonian.  The system exhibits exceptional points (EP) in the electronic spectrum at momenta 
\begin{align}
|\vec k_* | =  \frac{|\eta|}{\hbar u}, 
\end{align}
which are fully controlled by the non-Hermiticity of the system (real-valued parameter $\eta$). The EPs disappear when $\eta = 0$ as the system restores  full hermitian symmetry. 
In what follows below, we consider weak non-hermiticity ($|\eta| a/ \hbar u$$\ll$$1$, where $a$ is the lattice constant), and  
discuss the behavior of the \PT-symmetric Fermi liquid in the vicinity of the EP of non-hermitian Hamiltonian \footnote{We note however, that exactly at the EP, the non-hermitian Hamiltonian matrix is not diagonalizable \cite{Kato1966}}.

\textit{Fermi surface and Luttinger theorem for quantum systems with \PT symmetry.}---The Fermi liquid properties are defined by the shape and topology of its Fermi surface. The explicit real spectrum of the non-interacting problem allows us to extend the notion of the Fermi surface to the \PT-symmetric systems following. For this, we first "double" the Hamiltonian towards 
the full Hermitian \cite{doubled}; due to real eigenvalues of \eqref{Hamiltonian}, the doubled Hamiltonian has the Fermi surface of the same shape and topology as the original one \eqref{Hamiltonian}. For the purposes of our problem, the dynamics of $\mathcal H$ and $\mathcal H^{\dag}$ are effectively uncoupled.  
Furthermore, the introduction of the "doubled" Green's function \cite{doubled} allows defining  the matrix Green's function of the \PT-symmetric Hamiltonian \eqref{Hamiltonian} in the convenient form 
\begin{align}
\mathcal G (\vec k, \omega) = (\omega - \mathcal H (\vec k) ) ^{-1} . 
\end{align}
Upon diagonalization of \eqref{Hamiltonian}, the Fermi surface $\mathcal A_F$ is being defined through the Green's function as \cite{Abrikosov1963}
\begin{align}
 \lim _{\omega \to 0}  \mathcal G ^{- 1}  (\vec k , \omega)  |_{\mathcal A_F} \equiv 0. 
\end{align}
In particular, the Luttinger theorem holds \cite{Seki2017},  
\begin{align}
n = 2  \lim_{\omega \to 0} \int_{\mathcal G^{-1}  (\vec k, \omega) >0}  \frac{d^D \vec k}{(2 \pi)^D} . 
\label{Luttinger}
\end{align}
Note that definition above, while introduced for the noninteracting problem, are valid for the interacting FLs, provided the quasiparticle poles are well defined.

\textit{Electric conductivity of Fermi liquids.}---
Next, we revisit the electric conductivity of Fermi liquids, given by expression \cite{Lifshitz1971,Mahan2000} 
\begin{align}
 \sigma_{xx} = e^2 \sum_{\vec p,  s}  \left( - \frac{\partial f}{ \partial \varepsilon}   \right)  v^2_x(\vec p) \tau_{\vec p} ,
\end{align}
where $f (\eps)$ is Fermi-Dirac function, and $\tau_{\vec p}$ is the scattering time. In what follows below, we consider a uniform system with $\sigma_{xx}$$=$$\sigma_{yy}$$=$$\sigma_{zz}$$=$$
\sigma$$=$$\frac{1}{D} \Trace \sigma_{ij}$, $D$ is the dimensionality of the system. At $T=0$, $(-f' (\eps))$ behaves as a delta-function, hence we obtain
\begin{align}
 \sigma  = \frac{e^2}{D h^D} \lim_{\omega \to 0} \oint_{\mathcal G^{-1}  (\vec p, \omega) = 0}   d^{D-1} \vec p  \ |\vec v_{\vec p}| \tau_{\vec p} , 
 \label{Cond-int}
\end{align}
where we have used the property of the delta function $\delta[ g(x)] = \sum_i \frac{\delta(x- x_i)}{|g'(x_i)|}$, where $x_i$ are zeroes of $g(x)$.  Equation \eqref{Cond-int}, together with the Luttinger theorem \eqref{Luttinger}, defines the conductivity of a weakly-interacting Fermi liquid.

For the sanity check, we first rederive the \textit{Drude-Sommerfeld} conductivity \eqref{Drude} through Eqs. \eqref{Cond-int}, \eqref{Luttinger}. Consider a $D=3$ uniform Fermi gas with nonrelativistic dispersion relation $\eps = \hbar^2 k^2/2m$; it's Fermi surface (FS) is defined by \eqref{Luttinger} through relation $n = k_F^3/3 \pi^2$, where $p_F =\hbar k_F$ is Fermi momentum, and the quasiparticle velocity at FS is given by $v_F = \frac{1}{\hbar} \frac{\partial \eps}{\partial k} |_{k = k_F}$. Assuming momentum-independent scattering time, $ \tau_{\vec p}$$=$$\tau$, one arrives from Eq. \eqref{Cond-int} to Eq. \eqref{Drude}.

\begin{figure} [t]
\includegraphics[width = 0.85 \columnwidth]{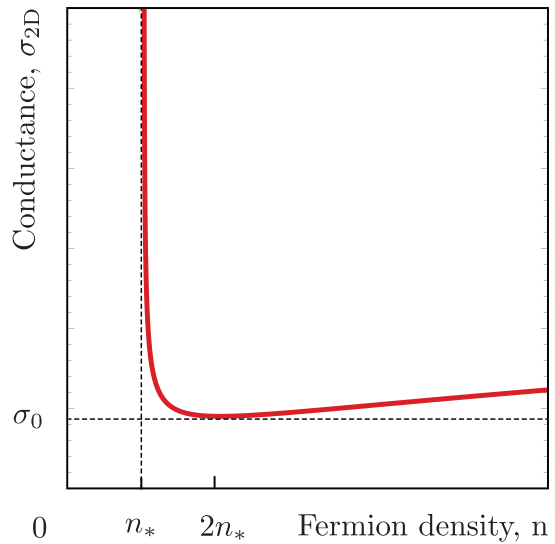}
\caption{Conductivity of  a \PT-symmetric Fermi liquid with Hamiltonian \eqref{Hamiltonian} as a function of electron density.    \label{Fig-sigma}}
\end{figure}

\textit{Conductivity of \PT-symmetric Fermi liquids.}---We now consider a 2D system with \PT-symmetry, defined in the low-energy approximation by Hamiltonian \eqref{Hamiltonian}, and  scattering time $ \tau_{\vec p}$$=$$\tau$. The expression Eqs. \eqref{Cond-int} in this case gives 
\begin{align} 
\sigma = \frac{e^2}{h} \frac{ u \tau k_F^2}{\sqrt{k_F^2 - k_*^2}}   .
\end{align}
Furthermore, using the Luttinger relation \eqref{Luttinger}, we find that the electronic conductance has the following scaling with electronic density $n$, 
\begin{align} 
\sigma = \frac{e^2}{\hbar} \frac{n u \tau}{\sqrt{2 \pi n- {\eta^2}/ { \hbar^2 u^2}}} . 
\end{align}
The dependece $\sigma (n)$ is illustrated in Figure \ref{Fig-sigma}.
We now discuss the properties of the conductivity. The first observation is that the conductivity is always metallic, with a minimum of $\sigma_0 = \frac{2 e^2}{\hbar} (k_* u \tau)$ at the density $n$$=$$2 n_*$. At large densities the metallicity improves, with scaling $\sigma (n$$\gg$$2 n_*) \sim \sqrt{n} \sim \varepsilon_F$, as expected since the effect of non-hermiticity is getting lost.    Notably, in the vicinity $n_*$$\le$$n$$\ll$$2 n_*$,  the system exhibits a low-resistance or a zero-resistance state (exactly at $n= n_*)$. This state is fully controlled by  the value of non-Hermiticity $\eta$ in the system's Hamiltonian \eqref{Hamiltonian}.

\textit{Zero-resistance states}.---Under range of validity of the present formalism,  zero resistance states  may appear on the brink of \PT-induced instability in the system by approaching the exceptional points  of the non-Hermitian Hamiltonian \eqref{Hamiltonian}. Detailed analysis shows that upon introducing the finite \PT perturbation $\eta$, the  system is  unstable at low densities $n$$<$$n_*$, however reclaims steady-state stability and well-defined Fermi-surface above  $n$$>$$n_*$, where $n_*$$=$$\eta^2 / 2 \pi \hbar^2 u^2$. Tuning density to $n=n_*$, the system undergoes through the exceptional points of the single-particle Hamiltonian,  the Fermi velocity diverges, 
\begin{align}
 |\vec v_{\vec k_F}|  \to \infty  \ \ \text{as}   \ \  k_F \to |\eta| /\hbar u.   
 \label{vF}
\end{align}
This, in return, boosts the electronic transport. Within this approach, the system experiences zero-resistance states (Fig. 2).  Away from the phase transition, $n$$\gg$$\eta^2 / 2 \pi \hbar^2 u^2$, the dense Fermi gas suppresses the effects of \PT-instability, and conductivity is smooth and anomaly-free, as expected for  Dirac fermions.

\begin{figure} [t]
\includegraphics[width = 1.0 \columnwidth]{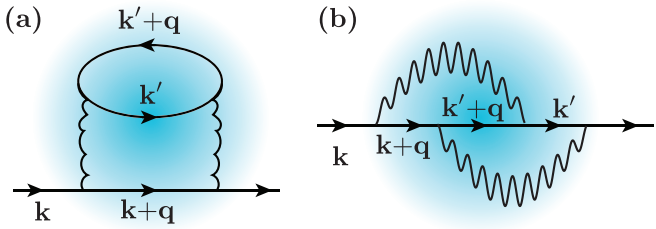}
\caption{Lowest order diagrams for electronic self-energy.  \label{diagram}}
\end{figure}

\textit{Violation of conventional temperature scalings.}---
Next, we show that the conventional temperature scalings of resistivity in  conventional  Fermi liquids, e.g. 
\begin{align}
    \rho_{\text{FL}}   \sim T^2 
    \label{Tsqrd}
\end{align} 
is violated in  \PT-symmetric Fermi liquid due to \PT anomalies in the single-particle spectrum \footnote{A generic interacting self-energy may contain higher powers in $T$. A typical argument distinguishing between a strange metal ($\rho_{\rm SM} \sim T$) and Fermi liquids involves temperature scaling at least as $ \rho_{\text{FL}}   \sim T^2$. }.
First, let us recall reasoning beyond \eqref{Tsqrd}, and then show how 
in the \PT-FLs this argument breaks down. To see the origin of this anomaly, let us revisit derivation of $\Sigma_{\vec k} (\omega, T)$ for Fermi liquids \footnote{Here we present a rather simplified analysis to show a typical $T$-scaling; full analysis of interacting self-energies can be found e.g. in Ref.\cite{Maslov2012b}.}.
A structured way to show this starts from considering the imaginary part of the finite-temeprature electronic self-energies  $\Sigma (\vec k,  i \omega_n)$,  with taking the DC limit ($\omega$$\to$$0$) in the end of calculations.   Consider the lowest order diagrams in Fig.\ref{diagram}, the first one gives
\begin{align}
   \Sigma_1 (\vec k,  i \omega_n) = - \frac{2 i}{\beta^2} \sum_{\vec q} U_{\vec q}^2   \sum_{\vec k'}\sum_{i \omega_0}  \sum_{i \omega'_m} 
   \mathcal G_{\vec k'} (i \omega'_m) 
   \nonumber 
   \\
    \times 
   \mathcal G_{\vec k + \vec q} (i \omega_n + i \omega_0) 
   \mathcal G_{\vec k' + \vec q} (i \omega'_m + i \omega_0) ,
   \nonumber
\end{align}
where $\omega_n$, $\omega'_m$ are fermionic and $\omega_0$ is bosonic Matsubara frequencies.  Upon analytic continuation, one obtains for the imaginary part of self-energy
\begin{align}
\Im  \Sigma^{R}_1 (\vec k,   \omega) =   - 2 \pi \sum_{\vec q} U_{\vec q}^2  \iint \limits_{-\infty}^{+\infty}  d w_0    d w'  \mathcal F(\omega, \omega'; \omega_0)  
\nonumber 
\\
\times  
A_{\vec k'} ( \omega')
A_{\vec k + \vec q} ( \omega + \omega_0)
A_{\vec k' + \vec q} ( \omega' + \omega_0),
\label{Sigma2}
\end{align}
where $A_{\vec k} ( \omega) = - \frac{1}{\pi} \Im  G^R_{\vec k} ( \omega)$ is the spectral function, and $G^R_{\vec k} ( \omega)$ is the causal (retarded) propagator, and 
\begin{align}
 \mathcal F(\omega, \omega'; \omega_0)  =   \left( f_+ (\omega') - f_+ (\omega'+ \omega_0) \right)
    \nonumber
    \\
     \times  \left( f_{-}(\omega_0) - f_{+} (\omega+ \omega_0)  \right) ,
\end{align}
where $f_{\pm}(\omega)$ are   Fermi-Dirac (-) and Bose-Einstein (+) distributions.

The following simplification comes for well-behaved (conventional) Fermi-liquids: since 
(i) $A_{\vec k} ( \omega)$ are sharp (behave as delta-functions), and (ii) Fermi velocity is finite and non-singular, the main contribution to  \eqref{Sigma2} comes from the terms where both initial and final states are on the Fermi surface $\mathcal A (\vec k)$. In this case, one can  formally set $\omega_0 \approx 0$ and $\omega' \approx 0$ in the propagators of the integrand \eqref{Sigma2}.
Thereafter, the $T$ dependence is encoded in function
\begin{align}
    \mathcal W ( \omega)  =   2 \iint \limits_{-\infty}^{+\infty}  d w_0    d w' \, 
     \mathcal F(\omega, \omega'; \omega_0) 
     =  \omega^2 + \pi^2 T^2, 
\end{align}
and the self-energy is factorized into the $\vec k$-dependent part and $\omega$-dependent part
\begin{align}
    \Im \Sigma (\vec k, \omega ) = C_{\vec k} \, \mathcal W (\omega ) = C_{\vec k} \, ( \omega^2 + \pi^2 T^2 ), 
    \label{Sigma3}
\end{align}
where $C_{\vec k}$ is momentum dependent, but not temperature dependent.  In particular, for conventional Fermi liquids, $C_{\vec k}$ has the following form \cite{Chubukov2012}
\begin{align}
    C^{(1)}_{\vec k} = 
    \frac{\pi}{2} \sum_{\vec q} U_{\vec q}^2 \oint_{\text{F.S.}}
    \frac{d \mathcal A_{\vec k_F}}{(2 \pi)^{D-1}}
    \frac{Z_{\vec k_F + \vec q} 
    Z_{\vec k'_F + \vec q}
    Z_{\vec k'_F }
    }{ | v_{\vec k'_F} |}
    \nonumber
    \\
    \times  \delta(\varepsilon_{\vec k_F + \vec q} ) \delta(\varepsilon_{\vec k'_F + \vec q} ) ,
    \label{C1}
\end{align}
for diagram 1, and
\begin{align}
    C^{(2)}_{\vec k} = 
   - \frac{\pi}{4} \sum_{\vec q} U_{\vec q}  \oint_{\text{F.S.}}
    \frac{d \mathcal A_{\vec k_F}}{(2 \pi)^{D-1}}
    \frac{Z_{\vec k_F + \vec q} 
    Z_{\vec k'_F + \vec q}
    Z_{\vec k'_F }
    }{ | v_{\vec k'_F} |}
    \nonumber
    \\
    \times  U_{\vec k'_F - \vec k_F}  \delta(\varepsilon_{\vec k_F + \vec q} ) \delta(\varepsilon_{\vec k'_F + \vec q} )  ,
    \label{C2}
\end{align}
for diagram 2. Here the $Z_{\vec k}$$\sim$$1$ is the quasiparticle weight and the integral $\oint_{\text{F.S.}}$ is taken over the $(D-1)$-dimensional Fermi surface (F.S.). 
Eliashberg \cite{Eliashberg1962} had proved that the expressions of type \eqref{Sigma3} holds to arbitrary order in perturbation theory of the interacting Fermi liquids after including higher-order diagrams. As a consequence, the 
 the electric resistivity of  conventional Fermi liquids scales at least as 
\begin{align}
   \rho_{\text{FL}} (T)  \propto \lim_{\omega \to 0}  \Im \Sigma  (\omega, T) \propto T^2. 
   \label{T^2}
\end{align}
 
We stress however that the transition from \eqref{Sigma2} to \eqref{Sigma3} does not hold in \PT-symmetric Fermi liquids. Indeed, the key assumption for the simplification of Eq. \eqref{Sigma2} was that both the initial state and final state remains on the Fermi surface. This is however violated in the zero-resistivity state of  \PT-FL since the Fermi velocity is singular, and the final state is by definition outside of the Fermi surface. Hence, one cannot set $\omega_0 =0$ and $\omega'= 0$ in the causal propagators of Eq. \eqref{Sigma2}, and  factorization in  \eqref{Sigma3} is not valid. Finally, one can also see from Eq.\eqref{C1}, \eqref{C2}, for which both initial and the final states are on the Fermi surface,  vanish (Fermi velocity $v_{\vec k}$ is in the denominator of expressions $C_{\vec k}$, while the numerator is $Z_{\vec k_F + \vec q} 
    Z_{\vec k'_F + \vec q}
    Z_{\vec k'_F} \sim 1$), hence the common structure of those diagrams at the Fermi surface gives \eqref{vF}
    \begin{align}
    \frac{Z_{\vec k_F + \vec q} 
    Z_{\vec k'_F + \vec q}
    Z_{\vec k'_F }
    }{ | v_{\vec k'_F} |} \to 0 \nonumber 
    \end{align}
    in \PT-FLs. 
    In other words, the expression \eqref{Sigma2} becomes non-perturbative in the vicinity of the zero-resistance anomaly  of \PT-FLs, and  the onvetional Fermi liquid scaling as in Eq. \eqref{T^2} is  violated. 

\vspace{3 mm} 

\textit{Discussion.}---We have considered electric conductivity of non-hermitian yet \PT-symmetric Fermi liquids with real-valued spectra  and find anomalies in the electric conductivity within the Fermi liquid formalism. These anomalies include  possible zero-resistivity states, in the vicinity of which the $T^2$ scaling law is violated.  Due to the nontrivial renormalization close to the EPs, the zero-resistance state may or may not preserve under inclusion of electron-electron interactions; we leave the exact analysis of the interacting ground state of \PT-FLs to future studies. Such analysis may be done within formalism of Ref. \cite{Pan2020}. 
Whether or not the superconducting instability may evolve from the zero-resistance instability of \PT-FLs is an open theory question; for a related discussion, we refer the Reader to Ref. \cite{Ghatak2018}. Importantly, non-Hermitian \PT-symmetric Hamiltonians of form \eqref{Hamiltonian} can be accessed experimentally on the basis of platforms  \cite{Jotzu2014} \& \cite{Kremer2019}, and the conductivity can be tested \textit{in situ} via  fluorescence microscopy \cite{Anderson2019}, including the $T$-dependence. We hope the results of this theory  can be accessed via measurements in future works.

\

\textit{Aknowledgements}. The author thanks Emil Bergholtz, Oleg Yazyev, Andrey Chubukov, Dmitrii Maslov, and Gregor Jotzu for useful discussions. This work was supported by the Branco Weiss Society in Science, ETH Zurich, through the grant on flat bands, strong interactions, and the SYK physics.

\bibliography{Refs}

\end{document}